\documentstyle[mprocl]{article}

\catcode`\@=11
\@addtoreset{equation}{section}
\catcode`\@=12
\begin{document}
\baselineskip8mm
\title{\vspace{-4cm}Simplest cosmological model with the scalar field}
\author{A. Yu. Kamenshchik$^{1 \dag}$, \ I. M. Khalatnikov$^{1,2
\dagger}$ and \ A. V. Toporensky$^{3 *}$}
\date{}
\maketitle
\hspace{-6mm}
$^{1}${\em L. D. Landau Institute for Theoretical Physics,
Russian Academy of Sciences, Kosygin str. 2, Moscow, 117334,
Russia}\\
$^{2}${\em Tel Aviv University,
Tel Aviv University, Raymond and Sackler
Faculty of Exact Sciences, School of Physics and Astronomy,
 Ramat Aviv, 69978, Israel}\\
$^{3}${\em Sternberg
Astronomical Institute, Moscow University, Moscow, 119899, Russia}\\
\\

We investigate the simplest cosmological model with the massive
real scalar non-interacting inflaton field minimally coupled to
gravity. The classification of trajectories in closed minisuperspace
Friedmann-Robertson-Walker model is presented.
The fractal nature of a set of infinitely bounced
trajectories is discussed.
The results of
numerical calculations are compared with those obtained by
perturbative analytical calculations around the exactly solvable
minisuperspace cosmological model with massless scalar field.
$^{\dag}$ Electronic mail:  kamen@landau.ac.ru\\
$^{\dagger}$ Electronic mail:  khalat@itp.ac.ru\\
$^{*}$ Electronic mail: lesha@sai.msu.su\\

\section{Introduction}
\hspace{\parindent}
In recent years cosmological models with a scalar field acquired the
great popularity because they can serve as a most natural basis for
the inflationary cosmology $^{1}$. Indeed, the presence of scalar
field in the model under consideration provides us with the existence
of effective cosmological constant at the early stage of the
cosmological evolution and with an opportunity to describe the decay
of this constant and the ``graceful exit'' from inflation and
transition to the Friedmann stage of evolution at a proper moment.

On the other hand scalar field is an integral part of modern
models in particle physics. Moreover, the main part of papers devoted
to quantum-cosmological description of the quantum origin of the
Universe and to the definition and construction of the wave function
of the Universe consider the models including scalar
field which after the ``birth'' of the Universe is driving
inflation $^{2-4}$.

Side by side with the comparatively simple models based on the simple
Lagrangians of the scalar field were developed rather complicated
schemes considering the non-minimally coupled scalar field $^{5}$,
complex scalar field $^{6,7}$ or the scalar field combining
complexity and non-minimal coupling $^{8,9}$.

However, even the dynamics of the simple cosmological model,
including gravity and minimally coupled scalar field with simple
potential including only massive term is rather rich and deserves
studying.  The dynamics of the minisuperspace cosmological models
with the massive real scalar field for the flat, open and closed
Friedmann universes was studied in papers $^{10,11}$ in terms of phase
space and theory of dynamical systems. It was noticed that the
dynamics of closed model (which is the most interesting from the
quantum-cosmological point of view) is more complicated than that
dynamics of open and flat models.  This dynamics allows the
transitions from expansion to contraction and the existence of points
of maximal expansion and minimal contraction in contrast with cases
of open and flat cosmologies.  Moreover, closed spherically symmetric
models cannot expand infinitely and should have the points of maximal
expansion provided the matter in the model under consideration
satisfies the condition of energodominance $^{12}$. The presence of
points of maximal expansion and minimal contraction open the
possibility for the existence of the trajectories of evolution of the
Universe escaping singularity and oscillating in a periodical $^{13}$
or in an aperiodical $^{14}$ way between turning points. The
possibility of existence of such trajectories or, in other words,
non-singular universes filled with scalar field was discussed also
earlier in Ref.  15.

Here we would like to consider the simplest cosmological model with
massive real minimally coupled scalar field without self-interaction.
We shall consider closed Friedmann model in minisuperspace including
only two variables -- cosmological radius $a$ and homogeneous mode of
inflaton scalar field $\varphi$. Because all the trajectories in
such a model have the point of maximal expansion one can give the
classification of the trajectories starting their evolution since
these points. Studying such a classification we ignore the
``prehistories'' of the trajectories under consideration, i.e. we do
not study their evolution before acchieving points of maximal
 expansion. Such an approach simplifies the classification, because
it gives us an opportunity to treat in the same way all the different
kinds of trajectories  -- trajectories which exist during some
finite intervals of time, i.e the trajectories which were born
in singularity and disappear in it, the infinitely
oscillating trajectories escaping falling into singularity and two
kinds of ``semi-infinite'' trajectories which were either born in the
singularity or disappear in it.

Suggested classification according to points of maximal expansion is
convenient because localization of these points (i.e. a set of
possible points of maximal expansion as well as those of minimal
contraction) is well known and was described in Ref.  10 in terms of
phase space and in Ref. 14 in terms of configuration space. General
formulas describing such points for more wide class of models were
presented in Ref. 9.

We shall distinguish trajectories which are monotonical in $a$ and
$\varphi$ i.e. trajectories which fall to singularity with
monotonically changing value of scalar field and the trajectories
possessing bounces i.e points of minimal contraction when $\dot{a}$
is equal to zero and trajectories having extrema in the value of
scalar field $\varphi$ -- we shall call them $\varphi$-turns. It is
remarkable that there is some regularity in the localization
of the points of maximal expansion corresponding to different kinds
of trajectories. The points is that the regions corresponding to
trajectories falling to singularity after some definite number of
$\varphi$-turns are separated by regions corresponding to bouncing
trajectories.
This phenomenon was detected by us due to numerical simulations
and will be presented in detail in the second section.

Third section will be devoted to comparison of our model with even
more simple model with massless scalar field $^{16}$. This model is
of interest because it is exactly solvable and have only trajectories
without bounces and $\varphi$-turns. So it is interesting to try
to design some kind of perturbation theory for our model with massive
scalar field where exactly solvable equations of motion for massless
model play role of zero-order approximation. Such scheme seems
promising because of interest to exactly solvable cosmological models
rising in recent years $^{17}$. We shall see that for our model one
can design such perturbative scheme and even to get the first-order
corrections in an explicit way. However, this perturbation theory
does not work in the vicinity of singularity. Nevertheless,
it will be shown that combining the solutions of equations of motion
obtained in the framework of perturbation theory with some general
properties of solutions of full (massive) theory one can obtain
some useful information concerning different regimes of cosmological
evolution. Perhaps, the approach developed in the present paper can
be useful for analysis of the dynamical behavior of cosmological
models which are close to some exactly solvable cases.

\section{The model, equations of motion and their numerical
investigation}
\hspace{\parindent}
We shall consider the cosmological model with an action
\begin{equation}
S = \int d^{4} x \sqrt{-g}\left\{\frac{m_{P}^{2}}{16\pi} R +
\frac{1}{2} g^{\mu\nu}\partial_{\mu}\varphi \partial_{\nu}\varphi
-\frac{1}{2}m^{2}\varphi^{2}\right\}.
\end{equation}
For the closed Friedmann model with the metric
\begin{equation}
ds^{2} = N^{2}(t) dt^{2} - a^{2}(t) d^{2} \Omega^{(3)},
\end{equation}
where
$a(t)$ is a cosmological radius, $N$ -- a lapse function and
$d^{2} \Omega^{(3)}$ is the metric of a unit 3-sphere and
with homogeneous scalar field $\varphi$
the action (2.1) takes the form
\begin{equation}
S = 2 \pi^{2} \int dt N a^{3}
\left\{\frac{3m_{P}^{2}}{8\pi}
\left[-\left(\frac{\dot{a}}{N a}\right)^{2} + \frac{1}{a^{2}}\right]
+\frac{\dot{\varphi}^{2}}{2 N^{2}} - \frac{m^{2}
\varphi^{2}}{2}\right\}.
\end{equation}

Now choosing the gauge $N = 1$ we can get the following equations of
motion
\begin{equation}
\frac{m_{P}^{2}}{16 \pi}\left(\ddot{a} + \frac{\dot{a}^{2}}{2 a}
+ \frac{1}{2 a} \right)
+\frac{a \dot{\varphi}^{2}}{8}
-\frac{m^{2} \varphi^{2} a}{8} = 0,
\end{equation}
\begin{equation}
\ddot{\varphi} + \frac{3 \dot{\varphi} \dot{a}}{a}
+ m^{2} \varphi = 0.
\end{equation}
Besides, we can write down the first integral of motion of our system
\begin{equation}
-\frac{3}{8 \pi} m_{P}^{2} (\dot{a}^{2} + 1)
+\frac{a^{2}}{2}\left(\dot{\varphi}^{2} + m^{2} \varphi^{2}\right)  =
0.
\end{equation}

It is easy to see from Eq. (2.6) that the points of maximal expansion
and those of minimal contraction, i.e. the points, where $\dot{a} =
0$ can exist only in the region where
\begin{equation}
\varphi^{2} \leq \frac{3}{4 \pi} \frac{m_{P}^{2}}{m^{2} a^{2}} ,
\end{equation}
which represents the field in the half-plane
$0 \leq a < +\infty, -\infty < \varphi <
+\infty$ restricted by hyperbolic curves $
\varphi \leq \sqrt{\frac{3}{4 \pi}} \frac{m_{P}}{m a}$
and
$\varphi \geq -\sqrt{\frac{3}{4 \pi}} \frac{m_{P}}{m a}$
(see Fig. 1).
Sometimes, region defined by nonequalities (2.7) is called Euclidean
or ``classically forbidden''. One can argue about validity of such
definition (see, for detail Ref. 9), but we shall use it for
brevity. Now we would like to distinguish between points of minimal
contraction where $\dot{a} = 0, \ddot{a} > 0$ and those of maximal
expansion where $\dot{a} = 0, \ddot{a} > 0$. Let us put
$\dot{a} = 0$, in this case one can express $\dot{\varphi}^{2}$ from
Eq.  (2.6) as
\begin{equation}
\dot{\varphi}^{2} = \frac{3}{4 \pi} \frac{m_{P}^{2}}{a^{2}}
-m^{2} \varphi^{2}.
\end{equation}
Substituting (2.8) and $\dot{a} = 0$ into Eq. (2.4) we have
\begin{equation}
\ddot{a} = \frac{4 \pi m^{2} \varphi^{2} a}{m_{P}^{2}}
-\frac{2}{a}.
\end{equation}
>From Eq. (2.9) one can easily see that the possible points of
maximal expansion are localized inside the region
\begin{equation}
\varphi^{2} \leq \frac{1}{2 \pi} \frac{m_{P}^{2}}{m^{2} a^{2}}
\end{equation}
while the possible points of minimal contraction lie outside this
region (2.10) being at the same time inside the Euclidean region
(2.7) (see Fig. 1).

We shall consider the trajectories beginning at some points of
maximal expansion in the region defined by (2.10).
As it has been already mentioned in the Introduction we shall trace
out only the parts of the trajctories  beginning since the points of
maximal expansion ignoring their ``prehistories''.

 For the velocities
$\dot{\varphi}$ defined by Eq.  (2.8) we choose direction ``up'',
i.e.  positive sign (the picture with velocities $\dot{\varphi}$
pointed out ``down'' can be obtained by mirror reflection in respect
to axis $\varphi = 0$).  Numerical investigation shows that for the
points of maximal expansion placed in the region denoted in Fig. 2 by
number {\bf 0} trajectories monotonically approach to singularity $a
= 0, \varphi = +\infty$ (some typical trajectories of this kind are
shown in Fig.  3a). It looks quite natural because in the region
close to the ordinate axis $a = 0$ behavior of our system is close to
that for the system with massless scalar field where bounces and
$\varphi$ - turns are absent.  Then trajectories beginning in the
region denoted in Fig. 2 by number {\bf 1} have a bounce. The typical
trajectories of this kind are shown in Fig. 3b.  Then we have region
which we shall denote by number $\bar{1}$.  In this region we also
have a bounce, but before this bounce we have $\varphi$ - turn. The
typical trajectories beginning in the region $\bar{1}$ are shown on
Fig. 3d.  The boundary between the regions {\bf 1} and $\bar{1}$ is
the place  where some periodical trajectories escaping singularity
can have their points of maximal expansion (such trajectories were
first mentioned in Ref. 13 and analyzed in detail in Ref. 14).  In
particular, through the point $\varphi = 0$ lying on the boundary
between the regions {\bf 1} and $\bar{1}$ goes the symmetric in
$\varphi$ periodic trajectory depicted in Fig.  3c.

In the region {\bf 1'} are localized the
trajectories which have not a bounce, but $\varphi$ - turn
after that they tend to singularity $a = 0, \varphi = - \infty$ (see
Fig. 3e).
The boundary between the regions $\bar{1}$
and {\bf 1'} corresponds to trajectories
where the points of bounce ($\dot{a} = 0, \ddot{a} > 0$)
degenerates into inflection points ($\dot{a} = 0, \ddot{a} = 0$).

Then in the region {\bf 2} (see Fig. 2) we have the points of the
maximal expansion for trajectories which after going through
$\varphi$-turn have a bounce at $\varphi < 0$ escaping in such a way
the fall into singularity  (see Fig. 3f).
In the region $\bar{2}$ we have the trajectories
which have a bounce in the lower half-plane, just after two
$\varphi$-turns. The boundary between the regions
{\bf 2} and $\bar{2}$ again contain trajectories
including turning points where we have simultaneously $\dot{a} = 0,
\dot{\varphi} = 0$. Some of these trajectories are periodical. One of
such trajectories is shown in Fig. 3g.

Then starting since the
point of maximal expansion in the region
{\bf 2'} (see Fig. 2) we have the trajectories
without bounces, but with two $\varphi$-turns (one at $\varphi > 0$
and one at $\varphi < 0$ ) after which they go to singularity $a = 0,
\varphi = +\infty$ (see Fig. 3h).

Now one can easily understand what
kind of trajectories corresponds to the points of maximal expansion
localized in regions {\bf 3}, $\bar{3}$,
{\bf 3'},
{\bf 4}, $\bar{4}$,
{\bf 4'} and so on.

It is worth noticing that the most peculiar form has the
boundary between the regions {\bf 1} and
 other regions in the upper half-plane ($\varphi > 0$). This
boundary consists from two curves (see Fig.2). The left one is finite
and divide the region {\bf 1} from the region
$\bar{1}$ while the right curve
is an infinite one and goes
almost parallelly to the hyperbolic curve separating the possible
points of maximal expansion from the possible points of minimal
contraction (see Fig.  2).
It is interesting that this upper curve touches all the regions
beginning since $\bar{1}$.

In Fig. 3i is shown the trajectory whose point of maximal
expansion is placed in the upper branch of the region {\bf 1}. This
trajectory immediately after the bounce has  $\varphi$ - turn, then
goes through the second point of maximal expansion falls into
singularity in the lower half-plane.

All the regions beginning since
$\bar{1}$ are compact in contrast with regions
{\bf 0} and {\bf 1}.

Now one can pay attention to a special class of trajectories having
the points of ``full stop'', i.e. the points where $\dot{a} = 0,
\dot{\varphi} = 0$ simulataneously. These points of full stop can lie
only on the boundary between Euclidean and Lorentzian region.
Naturally, the trajectories beginning with zero velocities at this
boundary constitute the tiny minority in the set of all possible
trajectories.  Nevertheless it is interesting to trace out the
correspondence between the points of full stops of these trajectories
and their points of maximal expansion. Here it is necessary to
remember that in every point of maximal expansion we have two
possible directions for $\dot{\varphi}$. Our classification was
constructed for the direction ``up''. The structure of regions
corresponding to the direction ``down'' can be obtained by reflection
of picture in respect to axis $\varphi = 0$. We shall distinguish
between regions ``up'' and ``down'' by symbols $\uparrow$ and
$\downarrow$ correspondingly.

Thus, let us consider the trajectories which begins with zero
velocities $\dot{a} = 0, \dot{\varphi} = 0$ in the upper half-plane
$\varphi > 0$. As was explained earlier the corresponding
trajectories can have their points of maximal expansion on the curves
separating the regions {\bf k} from the regions $\bar{k}$.  Besides
these boundaries the points of maximal expansion can be disposed on
the infinite curve separating the region {\bf 1} from all other
regions.  Indeed, numerical investigation shows that the trajectories
beginning at $0 < \varphi < \varphi_{0}$, go into Euclidean region
and have the first point of maximal expansion just on this infinite
curve separating the region {\bf 1}$\uparrow$ from other regions
Here $\varphi_{0}$ is the point where the direction of the
motion at the initial moment coincides with the direction of the
tangent to the curve given by an equality in (2.7) , i.e.  the point
where
\begin{equation}
\frac{\ddot{\varphi}}{\ddot{a}} = \frac{d
\varphi}{da}.
\end{equation}
Using Eqs. (2.4), (2.5),and (2.7) one
can find
\begin{equation}
\varphi_{0} = \sqrt{\frac{3}{4\pi}} m_{P}.
\end{equation}
Then we consider trajectories beginning with zero velocities
at $\varphi > \varphi_{0}$. Moving up the beginning of trajectory
along the hyperbolic curve separating Euclidean and Lorentzian
regions beginning since $\varphi_{0}$ we shall have the corresponding
point of maximal expansion moving down along the boundary between
the regions {\bf 1}$\uparrow$ and
$\bar{1}\uparrow$. Then, after approaching the
final point of this curve (i.e. the crossing between this boundary
curve and the curve separating the points of maximal expansion from
points of minimal contraction)
the closest to the beginning of motion point of maximal expansion
jumps to some point of the boundary between regions {\bf
2}$\downarrow$ and $\bar{2}\downarrow$. Then
while the beginning of the trajectory is moving up along the
hyperbolic curve separating Euclidean and Lorentzian regions the
corresponding point of maximal expansion is moving up to until the
moment when it approaches to the curve separating the points of
minimal contraction and maximal expansion in an upper half-plane.
Then it jumps to some point on the curve separating the regions
{\bf 3}$\uparrow$ and $\bar{3}\uparrow$ and
so on and so forth. Thus we have seen that every point on the upper
branch of the hyperbolic curve separating Euclidean and Lorentzian
regions has a counterpart on one of curves separating regions
{\bf k} and $\bar{k}$ and the intervals denoted
with sign ``$\uparrow$'' is interchanged with intervals denoted by
that of ``$\downarrow$''. The picture for the down part of hyperbolic
curve disposed at $\varphi < 0$ will be quite the same with an only
difference that signs $\uparrow$ should be substituted by ones
$\downarrow$ and vice versa.

Now let us turn back to the analysis of the general set of
trajectories in our problem.
The most interesting for us will be
trajectories which have bounces, and especially the trajectories
which have a lot of bounces and as an ideal situation an infinite
number of bounces (i.e.  the trajectories escaping singularity).

Let us look at the trajectories whose points of maximal expansion
are localized in the regions {\bf 1, 2, 3} \dots, or  in the regions
$\bar{1}$, $\bar{2}$,
$\bar{3}$, \dots
i.e. the trajectories having at least one bounce. We shall study the
structure of these regions from the point of view of localization of
their second point of maximal expansion.
The substructure of the region {\bf 1}$\uparrow$ is represented on
Fig.  4, while the forms of the trajectories corresponding to various
subregion of regions {\bf 1} and $\bar{1}$ are presented in Fig. 5.

On the right near the boundary with the region $\bar{1}$
we have subregion {\bf 1}$\uparrow$ $\bar{1}$ $\downarrow$ which
corresponds to the trajectories which have after a bounce the second
point of maximal expansion in the region $\bar{1}$ $\downarrow$ after
which they have the second bounce.  The next subregion
to the left from the subregion {\bf 1}$\uparrow$
$\bar{1}$$\downarrow$ is the subregion {\bf 1}$\uparrow$ {\bf
1'}$\downarrow$ which corresponds to the trajectories which after
bounce have the point of maximal expansion in the region {\bf
1'}$\downarrow$. After this point of maximal expansion they have a
$\varphi$ - turn and then fall to singularity $a = 0, \varphi =
-\infty$ .

Then we have the subregion  {\bf 1}$\uparrow$ {\bf 1}$\downarrow$
which corresponds to
trajectories with two bounces.
Then we have subregion {\bf 1}$\uparrow$ {\bf 1'}$\uparrow$.
Then follows the
subregions {\bf 1}$\uparrow$ {\bf 2}$\uparrow$,
{\bf 1}$\uparrow$ $\bar{2}$$\uparrow$,
{\bf 1}$\uparrow$ {\bf 2'}$\uparrow$,
{\bf 1}$\uparrow$ {\bf 1}$\uparrow$,
{\bf 1}$\uparrow$  {\bf 2'}$\downarrow$,
{\bf 1}$\uparrow$ {\bf 3}$\downarrow$ and so on and so forth.

The structure of region $\bar{1}$$\uparrow$
is much more simple: we have here only two subregions
$\bar{1}$$\uparrow$ {\bf 1}$\downarrow$ and
$\bar{1}$  {\bf 0}$\downarrow$.

Thus inside the regions {\bf 1}$\uparrow$,
$\bar{1}$$\uparrow$ we have an infinite set of
subregions whose arrangement approximately repeat the structure of
the whole field of the possible points of maximal expansion presented
in Fig.  2.

One can see that the structure of couples of regions
{\bf 2}$\bar{2}$; {\bf
3},$\bar{3}$; {\bf
4},$\bar{4}$ \dots is quite similar to that of
couple of regions {\bf 1},$\bar{1}$.

Studying the structure of subregions corresponding to two bounces
one can find inside each of them the same ``subsubstructure'' of
infinite set ``subsubregions''.
Continuing this procedure
{\it ad infinitum} one can observe that the field of the localization
of the points of maximal expansion corresponding to trajectories
escaping singularity can be find as a result of an infinite procedure
at each stage of each we encounter self-similar structures.
Such a self-similarity of structures appearing at different scales
points out on the fractal nature of the set obtained as the result
of infinite procedure $^{18}$. Thus, while the set of trajectories
escaping singularity and infinitely oscillating between points of
minimal contraction and maximal expansion has vanishing measure in
the set of all possible trajectories it can at the same time have
non-trivial fractal dimensionality. This phenomenon was first
discussed in a little bit different terms in the paper by Page
$^{14}$.

\section {The cosmological model with the massless scalar field
and the ``anti-slow-roll'' perturbation theory}
\hspace{\parindent}
We begin this paper with the rewriting of the equations (2.5) and
(2.6) in terms of conformal time
\[\eta = \int dt \frac{1}{a(t)}\]
because in the conformal time the solutions for the corresponding
equations of motion in massless limit can be written down explicitly.
Thus, we have
\begin{equation}
\varphi'' + \frac{2 \varphi' a'}{a} + m^{2} \varphi a^{2} = 0
\end{equation}
and
\begin{equation}
\frac{a'^{2}}{a^{2}} = -1 + \frac{4\pi}{3m_{P}^{2}} \varphi'^{2}
+\frac{4\pi}{3m_{P}^{2}} m^{2} \varphi^{2}.
\end{equation}
Here, ``prime'' denotes differentiation in respect to $\eta$.
At $m = 0$ Eqs. (3.1) and (3.2) turn into
\begin{equation}
\varphi'' + \frac{2 \varphi' a'}{a} = 0
\end{equation}
and
\begin{equation}
\frac{a'^{2}}{a^{2}} = -1 + \frac{4\pi}{3m_{P}^{2}} \varphi'^{2}.
\end{equation}
These system of equations (3.3) and (3.4) can be exactly integrated.
As in the preceding section we shall take as an initial moment of
evolution that at which the Universe is in the point of maximal
expansion. Then we can get
\begin{equation}
a = a_{0}\sqrt{\cos 2\eta},
\end{equation}
\begin{equation}
\frac{a'}{a} = -\tan 2\eta,
\end{equation}
\begin{equation}
\varphi' = \sqrt{\frac{3m_{P}^{2}}{4\pi}} \frac{1}{\cos 2\eta},
\end{equation}
(here, we have chosen direction ``up'' for the evolution of scalar
field just like in the preceding section)
and, finally
\begin{equation}
\varphi = \varphi_{0} + \frac{1}{4} \sqrt{\frac{3m_{P}^{2}}{4\pi}}
\ln \frac{1 + \sin 2\eta}{1 - \sin 2\eta}.
\end{equation}
These solutions correspond to trajectories which begin in one
singularity $a = 0, \varphi = \pm \infty$ go through the point of
maximal expansion $a = a_{0}, \varphi = \varphi_{0}$ and go to
other singularity $a = 0, \varphi = \mp \infty$ (see Ref. 16).
In terms of our Fig. 2 one can say that all half-plane $(a, \varphi)$
is filled with the only region {\bf 0}.

Now, our purpose is to construct the perturbation theory for the
solution of Eqs. (3.1) -- (3.2) using the solution of massless
equations (3.5)--(3.8) as zero-order approximation. Apparently, such
perturbation theory is predestined for describing situations when
the kinetic term in the Hamiltonian is much greater then the
potential term in contrast with the well-known slow-roll
approximation (see, for example, Ref. 7,19). Thus, can call our
scheme ``anti-slow-roll approximation''.
Again we choose as a starting moment for our evolution that moment,
when the Universe is placed in the point of maximal expansion. For
calculational simplicity we restrict ourselves by a symmetric case
$\varphi(0) = 0$. It will be more convenient to consider instead
of the variable $a$ the variable $h$ defined as
\begin{equation}
h \equiv \frac{a'}{a}.
\end{equation}

When representing the solutions of equations of motion as
\begin{equation}
\varphi = \varphi^{(0)} + \delta\varphi
\end{equation}
and
\begin{equation}
h = h^{(0)} + \delta h,
\end{equation}
where $\varphi^{(0)}$ and $h^{(0)}$ are given by formulae
(3.8) and (3.6) respectively, we can get the
following equations for $\delta\varphi$ and $\delta h$:
\begin{equation}
\delta\varphi'' + 2\delta\varphi' h^{(0)} + 2\delta
h \varphi^{(0)\prime} + m^{2} a^{(0)2} \varphi^{(0)}
\end{equation}
and
\begin{equation}
h^{(0)} \delta h =
\frac{4\pi}{3m_{P}^{2}}\varphi^{(0)\prime}\delta\varphi' +
\frac{4\pi}{3m_{P}^{2}} \frac{m^{2} \varphi^{(0)2} a^{(0) 2}}{2}.
\end{equation}
Substituting into Eqs. (3.12), (3.13) explicit expressions for
$\varphi^{(0)}, \varphi^{(0)\prime}, h^{(0)}$ and $a^{(0)}$
from Eqs. (3.5)--(3.8) we have
\begin{equation}
\delta\varphi'' - 2\tan 2\eta \delta\varphi'  +
2\delta h \sqrt{\frac{3m_{P}^{2}}{4\pi}} \frac{1}{\cos 2\eta}
+ \frac{1}{4}m^{2} a_{0}^{2} \cos 2\eta
\sqrt{\frac{3m_{P}^{2}}{4\pi}}
\ln \frac{1 + \sin 2\eta}{1 - \sin 2\eta}
\end{equation}
and
\begin{equation}
h^{(0)} \delta h =
\sqrt{\frac{4\pi}{3m_{P}^{2}}}
\frac{1}{\cos 2\eta}
\delta\varphi' +
\frac{m^{2}
a_{0}^{2}}{32} \ln^{2} \frac{1 + \sin 2\eta}{1 - \sin 2\eta}.
\end{equation}
One can easily see from Eqs. (3.14) and (3.15) that the true
small parameter in our perturbation theory is $m^{2}a_{0}^{2}$.

Now one can using Eq. (3.15) express $\delta h$ through
$\delta\varphi$. We have
\begin{equation}
\delta h =
- \sqrt{\frac{4\pi}{3m_{P}^{2}}}\frac{\delta\varphi'}{\sin 2\eta}
-\frac{m^{2}a_{0}^{2}}{32}\frac{\cos^{2} 2\eta}{\sin 2\eta}
\ln^{2} \frac{1 + \sin 2\eta}{1 - \sin 2\eta}.
\end{equation}
Substituting $\delta h$ from Eq. (3.16) into Eq. (3.14) we come to
the following equation for $\delta\varphi$:
\begin{eqnarray}
&&\delta\varphi'' - 2 \delta\varphi'
\left(\tan 2\eta + \frac{1}{\cos 2\eta \sin 2\eta}\right)\nonumber \\
&&-\sqrt{\frac{3m_{P}^{2}}{4\pi}} m^{2} a_{0}^{2}
\left(\frac{\cos 2\eta}{16 \sin 2\eta}
\ln^{2} \frac{1 + \sin 2\eta}{1 - \sin 2\eta}
-\frac{\cos 2\eta}{4}
\ln \frac{1 + \sin 2\eta}{1 - \sin 2\eta}\right) = 0.
\end{eqnarray}
This equation can be integrated and its solution looks as follows:
\begin{eqnarray}
&&\delta\varphi' = \sqrt{\frac{3m_{P}^{2}}{4\pi}}
\frac{m^{2} a_{0}^{2}}{16} \frac{\sin 2\eta}{\cos^{2} 2\eta}\times
\left\{2 \sin 2\eta\right.\nonumber \\
&&-\frac{1 + \sin^{2} 2\eta}{2 \sin 2\eta}
\ln^{2} \frac{1 + \sin 2\eta}{1 - \sin 2\eta}
-\cos^{2} 2\eta \ln \frac{1 + \sin 2\eta}{1 - \sin 2\eta}
+\ln^{2}(1-\sin 2\eta) - \ln^{2}(1+\sin 2\eta) \nonumber \\
&&\left.+ 2\ln 2 \ln \frac{1 + \sin 2\eta}{1 - \sin 2\eta}
+2 Li_{2} \left(\frac{1-\sin 2\eta}{2}\right)
-2 Li_{2} \left(\frac{1+\sin 2\eta}{2}\right) + C\right\}.
\end{eqnarray}
Here, in Eq. (3.18) $Li_{2}(x)$ is dilogarithm function, which can be
defined as a series $^{20}$:
\[Li_{2}(x) \equiv \sum_{n=1}^{\infty} \frac{x^{n}}{n^{2}}.\]
The constant of integration  $C$ in Eq. (3.18) should be chosen equal
to zero to provide the satisfaction of the condition
\[\delta h(0) = 0,\]
which means that the first-order correction to $h^{(0)}$
that can be obtained by substitution of Eq. (3.18) into Eq. (3.16)
should not
destroy the initial conditions for our evolution, i.e. that the
beginning of our consideration coincides with the point of maximal
expansion.

The behavior of  $\delta h$ and $\delta\varphi'$ are
presented in Fig.  6a and 6b respectively. We see that
$\delta\varphi'$ begins since zero value monotonically decreasing and
tending to $-\infty$ at $\eta \rightarrow \pi/4$, while $\delta h$
beginning since zero value monotonically increases and tends to
$+\infty$ at $\eta \rightarrow \pi/4$. Both first corrections have
opposite signs in respect with the corresponding functions in zero
approximation. So one can say that inclusion of the first corrections
$\delta\varphi'$ and $\delta h$ can describe the transition from the
set of trajectories without bounces and $\varphi$-turns to those,
having more complicated structure. It is interesting also to depict
the plots for the relative values
$\frac{\delta\varphi'}{\varphi^{(0)}}$ and $\frac{\delta h}{h^{(0)}}$
which are presented in Fig. 6c. The absolute value of
$\frac{\delta h}{h^{(0)}}$ is everywhere more than that of
$\frac{\delta\varphi'}{\varphi^{(0)}}$. It means that the we have
bounce earlier than $\varphi$-turn and in such a way our perturbative
scheme describes the transition from the region {\bf 0}
to the region {\bf 1} described in the preceding section. (It looks
quite natural that due to the first approximation of our perturbative
scheme we can describe only the small part of the general set of
trajectories, besides the regions disposed to the right from the
region {\bf 1} correspond to the comparatively large values of our
parameter $m^{2}a_{0}^{2}$ and hardly can be described by
perturbative methods).

However, here we should recognize that our
perturbative scheme fails in the vicinity of singularity. Indeed, it
is enough to look at the Fig. 6c to understand that $\frac{\delta
h}{h^{(0)}}$ reaches the value $-1$ at {\it any} value of the
parameter $m a_{0}$ that means that we shall have bounce on every
trajectory independently of the initial conditions. However, we know
that at small values of the radius of maximal expansion $a_{0}$ we
have the trajectories lying in the region {\bf 0}, i.e. the
trajectories without bounces and $\varphi$-turns. Thus, we see that
we can not, unfortunately, to estimate the parameters characterising
the change of the regimes of the behavior of our trajectories using
our perturbative scheme, because this scheme does not work in the
vicinity of singularity. However, one can try to extract some useful
information combining the perturbative results, numerical methods and
the equations describing the location of possible points of maximal
expansion and minimal contraction presented in the preceding section
 (see also Ref. 9).

First of all let us consider the symmetric trajectories, i.e
trajectories whose points of maximal expansion lies at
$\varphi = 0$. Numerical calculations show that the boundary between
the regions {\bf 0} and {\bf 1} lies at
\begin{equation}
m a_{0 \ \ {\rm numerical}} = 2.3.
\end{equation}
Now one can try to find this value using zero approximation of our
perturbative scheme or in other words, using the exact solutions
of the equations of massless model given by formulas (3.5)--(3.8).
Naturally, these equations describe the trajectories without bounces,
however, we can put forward the suggestion that zero-approximation
becomes invalid and the trajectories can have bounce when
these trajectories come into the region where points of minimal
contraction are localized, i.e. where according to Eq. (2.10)
\begin{equation}
\varphi a \geq \sqrt{\frac{1}{2\pi}} \frac{m_{P}}{m}.
\end{equation}
Substituting into Eq. (3.20) expressions for $\varphi(\eta)$ and
$a(\eta)$ from Eqs. (3.8) and (3.5) we have
\begin{equation}
m a_{0} \geq \frac{4 \sqrt{\frac{2}{3}}}
{\sqrt{\cos 2\eta} \ln \frac{1 + \sin 2\eta}{1 - \sin 2\eta}}.
\end{equation}
The value of $m a_{0}$ separating the regions {\bf 0} and {\bf 1} can
be found from the comparison of left-hand-side of non-equality (3.21)
with the maximal value of its right-hand-side. This value is as
follows:
\begin{equation}
m a_{0\ \ {\rm zero\  approximation}} = 1.6 .
\end{equation}

Now one can try to estimate the same value of the parameter $m a_{0}$
separating the regions {\bf 0} and {\bf 1} using the first
perturbative corrections to $a$ and $\varphi$. The logic of this
estimation will be in a way an opposite to that used in the
manipulation with zero-order (massless) solutions. Indeed, working
in zero-order approximation of the perturbation theory we have only
the trajectories without bounces. However, we know in that in some
region of our configuration space we can have bounces in the {\it
full} theory. Thus, fixing the moment when our ``zero-order''
trajectories come into this region we can estimate the value of the
parameter $m a_{0}$ giving the boundary between the regions {\bf 0}
and {\bf 1}. In first approximation {\it all} the trajectories have
a bounce at the moment when $\delta h = -h^{(0)}$, but we can again
estimate the value $\varphi a$ at the moment of bounce and to check
if it satisfies the non-equality (3.20). At some small values of
$m a_{0}$ this non-equality is broken and in such a way we can
estimate the boundary value of $m a_{0}$. Of course, these
calculations are much more complicated than those for zero-order
approximation and need some help of numerical methods.
It is interesting that the value of $m a_{0}$ parameterizing the
boundary between regions {\bf 0} and {\bf 1} is
\begin{equation}
m a_{0\ \ {\rm first\  approximation}} = 1.8
\end{equation}
and it is closer to the value obtained in the framework of full
theory (see Eq. (3.19)) than the value obtained in the framework
of zero-order approximation (see Eq. (3.22)).
Thus one can say that up to some extent our perturbation theory
complemented by some qualitative considerations describes the
transitions between different types of trajectories presented in
the full theory.
We would like to hope that the developed here perturbative scheme
can be useful in other problems where we have opportunity to consider
the exactly solvable model which is in some kinship with the model
under consideration.\\
\\
{\bf ACKNOWLEDGMENTS}

We are grateful to
D.N. Page and V. Sahni for useful discussions. This work was supported by
Russian Foundation for Fundamental Researches via grants
No 96-02-16220 and No 96-02-17591
and by INTAS via project 93-3364-Ext.
I.M.K. is grateful to
School of Physics and Astronomy   Tel Aviv University for constant
support.

\begin{description}
\item[\rm 1.] A.D. Linde,
{\it Particle Physics and Inflationary Cosmology} (Harwood Academic,
1990) and references therein.
\item[\rm 2.] J.B. Hartle and S.W.
Hawking, {\it Phys. Rev.} {\bf D28}, 2960 (1983); S.W. Hawking, {\it
Nucl.  Phys.} {\bf B239}, 257 (1984).
\item[\rm 3.] A. Vilenkin,
{\it Phys.  Lett.} {\bf 117B}, 25 (1982); {\it Phys.  Rev.} {\bf
D27}, 2848 (1983); {\it Phys. Rev.} {\bf D30} 509 (1984); {\it Phys.
Rev.} {\bf D37}, 888 (1988); A.D. Linde, {\it Zh. Eksp.  Teor.  Fiz.}
{\bf 87}, 369 (1984)  [{\it Sov. Phys. JETP} {\bf 60}, 211 (1984)];
Ya.B.  Zeldovich and A.A. Starobinsky, {\it Pis'ma Astron.  Zh.}
{\bf 10}, 323 (1984) [{\it Sov.  Astron. Lett.} {\bf 10}, 135
(1984)]; V.A.  Rubakov, {\it Phys. Lett.} {\bf 148B}, 280 (1984).
\item[\rm 4.] A.O. Barvinsky, {\it Phys. Rep.} {\bf 230}, 237 (1993);
A.O. Barvinsky and
A.Yu. Kamenshchik, {\it Phys. Rev.}  {\bf D50}, 5093 (1994).
\item[\rm 5.] B.L.Spokoiny, {\it
Phys.  Lett.}  {\bf 129B}, 39 (1984); D.S. Salopek, J.R. Bond and J.M.
Bardeen, {\it Phys.  Rev.} {\bf D40}, 1753 (1989); R. Fakir and W.G.
Unruh, {\it Phys.  Rev.}  {\bf D41}, 1783 (1990); R. Fakir, S. Habib
and W.G.  Unruh, {\it Aph.  J.}  {\bf 394}, 396 (1992); R. Fakir and
S. Habib, {\it Mod.  Phys. Lett.}  {\bf A8}, 2827 (1993).
\item[\rm 6.] I.M. Khalatnikov and A. Mezhlumian, {\it Phys.
Lett.} {\bf A169}, 308 (1992); I.M. Khalatnikov and P. Schiller,
{\it Phys. Lett.}  {\bf B302}, 176 (1993).
\item[\rm 7.]
L. Amendola, I.M. Khalatnikov, M. Litterio and F.  Occhionero,
{\it Phys. Rev.}  {\bf D49},
1881 (1994).
\item[\rm 8.] A.Yu.
Kamenshchik, I.M.  Khalatnikov and A.V.  Toporensky, {\it Phys. Lett.}
{\bf B357}, 36 (1995).
\item[\rm 9.] A.Yu.
Kamenshchik, I.M.  Khalatnikov and A.V.  Toporensky,
{\it Int. J. Mod. Phys.} {\bf D} (in press) // gr-qc/9801039.
\item[\rm 10.] V.A. Belinsky, L.P. Grishchuk,
Ya.B. Zel'dovich and I.M.  Khalatnikov, {\it J. Exp. Theor. Phys.}
{\bf 89}, 346 (1985).
\item[\rm 11.]
V.A.  Belinsky and I.M. Khalatnikov, {\it J.
Exp.  Theor.  Phys.}  {\bf 93}, 784 (1987);
V.A.  Belinsky, H. Ishihara, I.M. Khalatnikov and H. Sato,
{\it Progr. Theor. Phys.} {\bf 79}, 676 (1988).
\item[\rm 12.] G.A. Burnett, {\it Phys. Rev.} {\bf D51}, 1621 (1995).
\item[\rm 13.] S.W. Hawking, in {\it Relativity, Groups and Topology
 II}, ed. B.S. DeWitt and R. Stora, (North Holland, Amsterdam, 1984).
\item[\rm 14.] D.N. Page, {\it Class. Quantum Grav.} {\bf
1}, 417 (1984).
\item[\rm 15.] L. Parker and S.A. Fulling, {\it Phys. Rev.} {\bf D7},
2357 (1973); A.A. Starobinsky, {\it Pisma A.J.} {\bf 4} 155 (1978).
\item[\rm 16.] V.A.  Belinsky and I.M. Khalatnikov, {\it J.
Exp.  Theor.  Phys.}  {\bf 63}, 1121 (1972).
\item[\rm 17.]
R. de Ritis, G. Marmo, G. Platania, C. Rubano, P. Scudellaro
and C. Stornaiolo, {\it Phys. Rev.} {\bf D42}, 1091 (1990);
{\it Phys. Lett.} {\bf 149A}, 79 (1990);
S. Capozziello, R. de Ritis and P. Scudellaro, {\it Int. J.
Mod. Phys.} {\bf D3}, 609 (1994);
J.D. Barrow, {\it Phys. Rev.} {\bf D49}, 3055 (1994);
F.E. Schunk and E.W. Mielke, {\it Phys. Rev.} {\bf D50}, 4794 (1994);
S. Capozziello, M. Demianski, R. de Ritis and C. Rubano, {\it Phys.
Rev.} {\bf D52}, 3288 (1995).
\item[\rm 18.]
B. Mandelbrot, {\it The fractal geometry of nature}, (San Francisco,
Freeman, 1982).
\item[\rm 19.]
G.W. Lyons, {\it Phys. Rev.} {\bf D46}, 1546 (1992);
R. Bousso and S.W. Hawking,
{\it Phys. Rev.} {\bf D52}, 5659 (1995);
A.O. Barvinsky, A.Yu. Kamenshchik and I.V. Mishakov,
{\it Nucl. Phys.} {\bf B491}, 387 (1997).
\item[\rm 20.]
M. Abramowitz and I.A. Stegun, {\it Handbook of mathematical
functions}, (New York, Dover, 1972).
\end{description}

\newpage
\begin{center}
{\bf Captions to Figures}
\end{center}

{\bf Fig. 1.}  On the half-plane $a \geq 0,\varphi$ the solid
hyperbolic curves are those separating Lorentzian region from
Euclidean one. Dashed hyperbolic curves separate the possible points
of maximal expansion $\dot{a} = 0, \ddot{a} < 0$ (these points can
exist between two branches of dashed curves) and the possible points
of minimal contraction $\dot{a} = 0, \ddot{a} > 0$ (these points can
exist between the dashed and solid hyperbolic curves).

{\bf Fig. 2.} The structure of the region of the localization of
possible points of maximal expansion is presented. We consider here
the points of maximal expansion corresponding to the trajectories
which at these points has the direction ``up'', i.e. $\dot{\varphi} >
0$.

The region {\bf 0} corresponds to the trajectories which after
going through the point of maximal expansion placed in this region
go to the singularity $a = 0, \varphi = +\infty$.

The region {\bf 1}
corresponds to the trajectories which after the going through the
points of maximal expansion have a point of minimal contraction or
``bounce''.

The region {\bf 1'} corresponds to the trajectories which
after the going through the point of maximal expansion have a
$\varphi$-turn, i.e. the point where $\dot{\varphi} = 0$ after that
they fall to singularity $a = 0, \varphi = -\infty$.

The region {\bf
2} corresponds to the trajectories which after the going through the
point of maximal expansion and subsequent $\varphi$ - turn have
bounce.

The region {\bf 2'} corresponds to the trajectories which
after going through the point of maximal expansion and two subsequent
$\varphi$ - turns fall to singularity and so on.

We did not depicted
the regions $\bar{1}, \bar{2}, \dots$ described in the main text.
These regions are included into regions {\bf 1',2', \dots}
correspondingly.

{\bf Fig. 3.} The parts of the trajectories corresponding to the
regions presented on the preceding Figure are depicted. We trace out
for the sake of simplicity of the classification only the pieces of
these trajectories beginning since the point of maximal expansion
ignoring their ``prehistories''. On the {\bf Fig. 3a -- 3h} we
consider the trajectories whose points of maximal expansion are
placed at the line $\varphi = 0$, though it is not essential from the
general point of view.

In {\bf Fig. 3a} is presented the trajectory
corresponding to the region {\bf 0} falling after the point of
maximal expansion to the singularity.

In {\bf Fig. 3b} the trajectory
corresponding to the region {\bf 1} is presented. This trajectory has
a bounce after the going through the point of maximal expansion.

In
{\bf Fig. 3c} is presented the periodic trajectory having two
symmetric points of ``full stops'' $\dot{a} = 0, \dot{\varphi} = 0$
at the boundary between Lorentzian and Euclidean regions and the
point of maximal expansion placed just at the boundary between the
regions {\bf 1} and {\bf 1'} (or, to be precise on the boundary
between the regions {\bf 1} and $\bar{1}$.

In {\bf Fig. 3d} the trajectory corresponding to the region $\bar{1}$
is presented. This trajectory after the point of maximal expansion
has $\varphi$ - turn, almost immediately after it -- bounce, then the
second point of maximal expansion and then it goes to singularity.

In {\bf Fig. 3e} we have the trajectory corresponding to the region
{\bf 1'} which after $\varphi$ - turn falls to infinity.

In {\bf Fig. 3f} we see the trajectory corresponding to the region
{\bf 2} which after the $\varphi$ - turn has a bounce already in the
lower half-plane.

In {\bf Fig. 3g} we see the periodic trajectory whose point of
maximal expansion lies on the axes $\varphi = 0$ just at the boundary
between the regions {\bf 2} and $\bar{2}$.

In {\bf Fig. 3h} is shown the trajectory corresponding to the region
{\bf 2'} which after two $\varphi$ - turns fall to upper singularity.

In {\bf Fig. 3i} is shown the trajectory whose point of maximal
expansion is placed in the upper branch of the region {\bf 1}. This
trajectory immediately after the bounce has  $\varphi$ - turn, then
goes through the second point of maximal expansion falls into
singularity in the lower half-plane.

{\bf Fig. 4.}  The structure of the subregions of the region {\bf 1}
is presented. Thin regions denoted by letters ``b,c,d,e'' correspond
to the trajectories which have at least two bounces. The subregion
``a'' and other subregions placed between thin regions correspond to
the trajectories falling to singularity after one bounce.

{\bf Fig. 5.} The trajectories corresponding to different subregions
of the region {\bf 1} are presented. The trajectories depicted on the
{\bf Fig. 5a -- 5e} correspond respectively to the subregions
``a--e'' presented in {\bf Fig. 4}. The trajectory depicted on {\bf
Fig. 5a} has only one bounce while the trajectories have at least two
bounces. The trajectories shown in {\bf Fig. 5c} and {\bf Fig. 5e}
have a point of full stop (or at least are very close to it).

{\bf Fig. 6.} In {\bf Fig. 6a} is presented the dependence of the
$\delta h$ on the conformal time $\eta$ while in {\bf Fig. 6b} the
dependence of $\delta\varphi'$ on $\eta$ is presented. In {\bf Fig.
6c} solid line shows the behavior of $\delta h /h$ and dashed line
shows the behavior of $\delta\varphi' / \varphi'$.

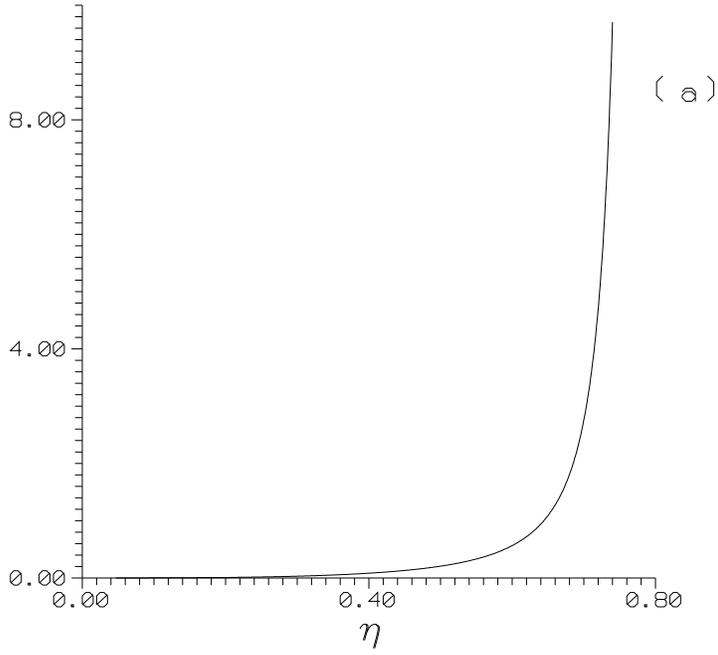
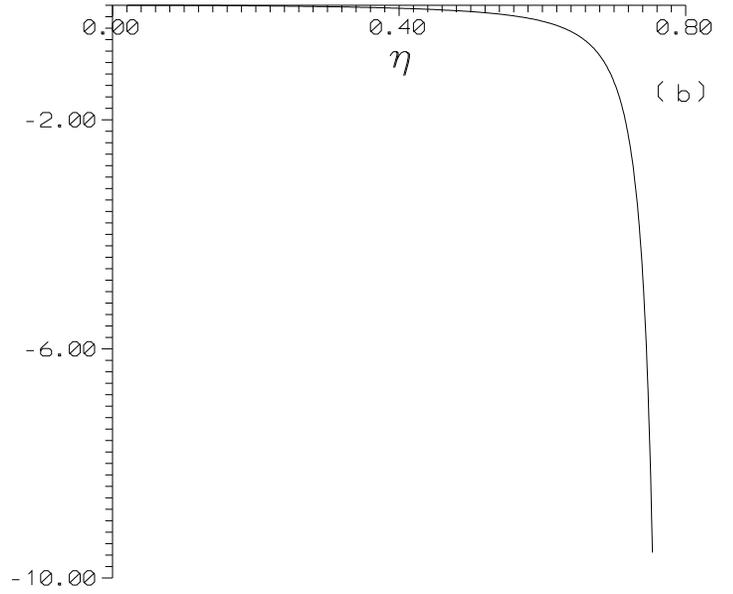
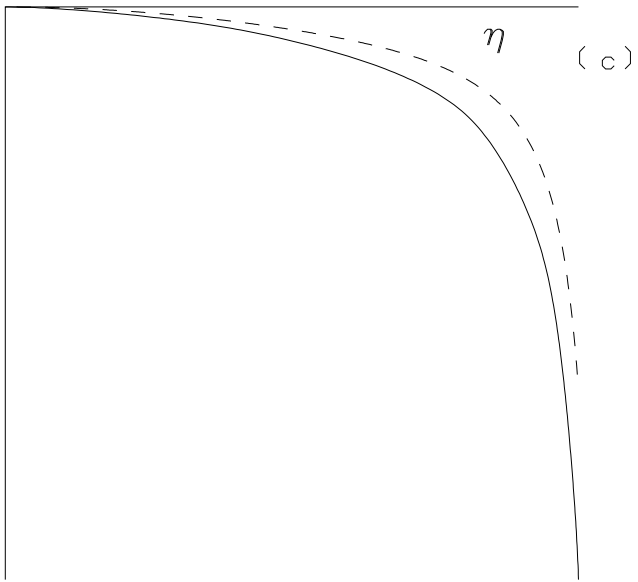
\end{document}